\documentclass[preprint,aps,amsmath,superscriptaddress,nofootinbib]{revtex4}
\usepackage[a4paper, margin=0.79 in]{geometry}
\usepackage{graphicx}  %
\usepackage{amsmath}
\usepackage{slashed}
\usepackage{bm}  %
\usepackage{ulem}
\usepackage{mwe}
\usepackage{verbatim}
\usepackage[titletoc,title]{appendix}
\usepackage{comment}
\usepackage{wrapfig}
\usepackage{lipsum}

\def\Dsl{\hbox{/\kern-.6000em D}} 

\def\dsl{\,\raise.15ex\hbox{/}\mkern-13.5mu D}

\def\psip#1{\psi_{\mathbf{#1}}}

\def\ltap{\ \raise.3ex\hbox{$<$\kern-.75em\lower1ex\hbox{$\sim$}}\ }
\def\gtap{\ \raise.3ex\hbox{$>$\kern-.75em\lower1ex\hbox{$\sim$}}\ }

\def\OMIT#1{}
\newcommand{\nn}{\nonumber}

\newcommand{\bea}{\begin{eqnarray}}
\newcommand{\eea}{\end{eqnarray}}

\def\mqo2{{\!\!\!}}

\begin{document}
\title{
Perturbative Corrections to Heavy Quark-Diquark Symmetry Predictions for Doubly Heavy Baryon Hyperfine Splittings}
\author{Thomas Mehen}
\email{mehen@phy.duke.edu}
\affiliation{Department of Physics,
         Duke University, Durham, NC\ 27705, USA\\}

\author{Abhishek Mohapatra}
\email{abhishek.mohapatra@duke.edu}
\affiliation{Department of Physics,
         Duke University, Durham, NC\ 27705, USA\\}

\date{\today}
\begin{abstract}
Doubly heavy baryons $\left(QQq\right)$ and singly heavy antimesons $\left(\bar{Q}q\right)$ are related by the heavy quark-diquark (HQDQ) symmetry because in the $m_Q \to \infty$ limit, the light degrees of freedom in both the hadrons are expected to be in identical  configurations. Hyperfine splittings of the ground states in both systems are nonvanishing at $O(1/m_Q)$ in the heavy quark mass expansion and HQDQ symmetry relates the hyperfine splittings in the two sectors.   In this paper, working within the framework of Non-Relativistic QCD (NRQCD), we point out the existence of an operator that couples four heavy quark fields to the chromomagnetic field with a coefficient that is enhanced by a factor from Coulomb exchange. This operator gives a correction to doubly heavy baryon hyperfine splittings that scales as $1/m_Q^2 \times \alpha_S/r$, where $r$ is the separation between the heavy quarks in the diquark. This correction  can be calculated analytically in the extreme heavy quark limit in which the potential between the quarks in the diquark is  Coulombic. In this limit, the correction is $O(\alpha_s^2/m_Q)$ and comes with a small coefficient. For values of $\alpha_s$ relevant to doubly charm and doubly bottom systems, the correction to the hyperfine splittings in doubly heavy baryons is only a few percent or smaller. We also argue that nonperturbative corrections to the prediction for the hyperfine splittings are suppressed by $\Lambda^2_{\rm QCD}/m_Q^2$ rather than $\Lambda_{\rm QCD}/m_Q$. Corrections should be $\approx 10\%$ in the charm sector and smaller in heavier systems.
\end{abstract}
\maketitle
\newpage
\section{Introduction}

The first doubly charm baryon $\Xi_{cc}^{++}$ with mass $3621.40\pm 0.72\pm0.27\pm 0.14$ MeV was recently observed by the LHCb collaboration in the exclusive decay modes, $\Xi_{cc}^{++}\rightarrow \Lambda_c^{+}K^{-}\pi^{+}\pi^{+}$ and $\Xi_{cc}^{++}\rightarrow \Xi_c^{+}\pi^{+}$ \cite{Aaij:2017ueg, Aaij:2018ueg}. Even though the SELEX collaboration 
\cite{Mattson:2002vu, Moinester:2002uw, Ocherashvili:2004hi} had earlier reported the observation of doubly charmed baryons years ago, those observations were not confirmed by other experiments such as FOCUS \cite{Ratti:2003}, Belle \cite{Chistov:2006, Kato:2013ynr}, and BaBar \cite{Aubert:2006}. The large isospin violation implied by the recent LHCb results also cast doubt on the validity of the SELEX results. The recent experimental observation of the $\Xi_{cc}^{++}$ baryon has greatly revived the interest in the physics of doubly heavy baryons. This includes the experimental efforts to search for the other doubly charm and bottom baryons such as $\Xi_{cc}^+$ and $\Xi_{bc}$ \cite{Traill:2017zbs} as well as recent theoretical studies regarding the lifetimes, production rates, and decay rates of the double heavy baryons~\cite{Cheng:2018,Berezhnoy:2018, Cheng:2019, Karliner:2014, Kiselev:2001fw, Ridgway:2019}.

An interesting idea regarding the physics of doubly heavy baryons is that of heavy quark-diquark (HQDQ) symmetry  which relates the physics of doubly heavy baryons $\left(QQq\right)$ to the heavy antimesons $\left(\bar{Q}q\right)$. The appropriate theory for dealing with heavy mesons is the  heavy quark effective field theory (HQET)~\cite{Georgi:1990ak,Falk:1990yz,Eichten:1989zv}, whereas the appropriate theory for dealing with doubly heavy baryons is  nonrelativistic quantum chromodynamics (NRQCD) \cite{Bodwin:1994jh,Luke:1999kz, Brambilla:1999}. In the limit of large heavy quark mass, $m_Q$, the two heavy quarks, $QQ$, in the doubly heavy baryon experience an attractive Coulomb force and the ground state of the two heavy quarks is a tightly bound spin-$1$ diquark in the $\bar{3}$ color representation. The size of the diquark is small, $r\sim \left(1/m_Qv\right)^{-1}\ll\Lambda_{\rm QCD}^{-1}$, where $v$ is the relative velocity of the two heavy quarks in the diquark. This implies that the diquark can be considered as a point source of color charge in the $\bar{3}$ representation that looks the same to the light degrees of freedom as a singly heavy antiquark, up to corrections that are suppressed by inverse powers of heavy quark mass $m_Q$. The light degrees of freedom in the heavy antimeson also orbit a point source of color charge in the $\bar{3}$ representation. Therefore, the two heavy hadrons have identical configurations for the light degrees of freedom in the $m_Q\to \infty$ limit. The HQDQ symmetry also relates the double heavy tetraquarks to singly heavy baryons and the chiral lagrangians incorporating this symmetry have been derived in Refs.~\cite{Hu:2005gf,Mehen:2017nrh}.

One of the implications of the HQDQ symmetry is the relation between the hyperfine mass splittings of the doubly heavy baryons and heavy antimesons. The chromomagnetic interactions of the diquark and quark are responsible for the hyperfine splittings in the doubly heavy baryons and antimesons. The effective Lagrangian describing the chromomagnetic coupling of diquarks at  
${\cal O}\left(1/m_Q\right)$ was derived in Ref.~\cite{Mehen:2006,Brambilla:2005} in the framework of NRQCD. The ground state of the heavy antimeson consists of a spin-$0$ meson, $P$, and a spin-$1$ meson, $P^*$. The ground state of the doubly heavy baryon consists of a spin-$1/2$ baryon, $\Xi$, and a spin-$3/2$ baryon $, \Xi^*$. These states are degenerate due to heavy quark spin symmetry that breaks at ${\cal O}\left(1/m_Q\right)$ due to spin dependent chromomagnetic interactions. The heavy quark-diquark symmetry implies the relation between the hyperfine splittings to be  \cite{Savage:1990di,Mehen:2006,Brambilla:2005}
\begin{equation}
    m_{\Xi^*}-m_{\Xi}=\frac{3}{4}\left(m_{P^*}-m_{P}\right).
    \label{hyperfine:HQDQ}
\end{equation}

The purpose of this paper is to study higher order corrections to this prediction. Since the hyperfine splittings themselves are $O(1/m_Q)$, one might expect the leading corrections
to scale as $1/m_Q^2$. What we will see below is that there is a higher dimension operator that scales as $1/m_Q^2 \times \alpha_s/r$, where $r$ is the typical separation between the quarks within the diquark. Since $1/r \sim m_Q v$, this operator gives a correction to Eq.~(\ref{hyperfine:HQDQ}) of relative order $\alpha_s v$. In the extreme limit where the quarks within the diquark are bound by Coulombic gluon exchange, $v$ is proportional to $\alpha_s$ and this $O(\alpha_s^2)$ correction is computed below. 
$O(\alpha_s^2)$ corrections to the prediction for the doubly charm hyperfine splittings were first anticipated in Eq.~(35) of Ref.~\cite{Brambilla:2005}. Here we compute this correction explicitly for the first time, and it turns out to be only of order a percent or less for values of $\alpha_s$ relevant to doubly charm and bottom baryons. A similar correction to the HQDQ symmetry due to the finite size of diquark was also calculated in Ref.~\cite{ An:2019}. The finite size effects were due to operators coupling the light quarks and the diquarks that contribute to the mass of the double heavy baryon. The correction to HQDQ symmetry was also estimated to be small in Ref.~\cite{An:2019}. In this paper, we also argue that the leading corrections to the prediction in Eq.~\eqref{hyperfine:HQDQ} will be the nonperturbative corrections to both the hyperfine splittings scaling as $O(\Lambda_{\rm QCD}^2/m_Q^2)$, which has not yet been computed. In the body of this paper we review the effective action for heavy diquark fields, introduce the  operator and compute its effect on the prediction for doubly heavy baryon hyperfine splittings. We then give our argument for why the nonperturbative corrections to the prediction for the hyperfine splitting in Eq.~\eqref{hyperfine:HQDQ}  are suppressed by $\Lambda_{\rm QCD}^2/m_Q^2$.
This is followed by our conclusions. In an Appendix, we derive the form of the operator by matching the full QCD diagrams for $QQ g\to QQ$ scattering onto NRQCD to $O(1/m_Q^2).$

\section{Effective action for Composite diquark fields}
The effective action for the heavy composite diquark fields with the lowest order heavy quark spin symmetry violating  chromomagnetic interaction was derived by  Fleming and Mehen in Ref.~\cite{Mehen:2006} and Brambilla, Vairo, and Rosch in Ref.~\cite{Brambilla:2005} in the framework of NRQCD. The leading order chromomagnetic couplings of diquarks gives ${\cal O}\left(1/m_{\rm Q}\right)$ corrections to the heavy quark spin symmetry and is responsible for the hyperfine splittings in the ground state of doubly heavy baryons. In this section, we use the formalism  in Ref.~\cite{Mehen:2006} to include the correction to the chromomagnetic coupling of diquark fields from diagrams that contribute to the effective action  at higher order in NRQCD power counting.
 
The NRQCD Lagrangian relevant for constructing the effective action for composite diquarks is 
\begin{eqnarray}\label{nrqcd}
{\cal L} &=&  
 -\frac{1}{4}F^{\mu\nu}F_{\mu \nu} + 
\sum_{\bf p} \psip p ^\dagger   \Biggl ( i D^0 - \frac{\left({\bf p}-i{\bf D}\right)^2}{2 m_Q} 
 + \frac{g}{2 m_Q} \,\bm{\sigma}\cdot {{\bf B}} 
 \Biggr )
 \psip p  \nonumber \\
&& - \frac{1}{2} \sum_{\bf p,q}
 \frac{g_s^2}{ \bf (p-q)^2} 
  \psip q ^\dagger  T^A \psip p \psip {-q}^\dagger  T^A \psip {-p} + \ldots \, ,
\end{eqnarray}
 where $\psip p$ represents the quark field with a three vector label ${\bm p}$, ${\bm B}$ is the chromomagnetic field, and the ellipsis represents the higher order corrections as well as terms including soft gluons. 
The color and spin Fierz identities that project the potential into color anti-triplet $\left(\bar 3\right)$ and color sextet $\left(6\right)$ states and decompose the quark billinears such as $\psip p\psip {-p}$ into operators of definite spin are
\begin{align}
&\delta_{\alpha \delta} \delta_{\beta \gamma} = -\frac{1}{2}(\sigma^i \epsilon)_{\alpha \beta} 
(\epsilon \sigma^i)_{\gamma \delta}  + \frac{1}{2} \epsilon_{\alpha \beta} \epsilon_{\delta \gamma}, \label{spin} \\
&T^a_{il}T^a_{jk} = -\frac{2}{3}\sum_m \frac{1}{2}\epsilon_{mij} \epsilon_{mlk}
+\frac{1}{3}\sum_{(mn)} d^{\,(mn)}_{ij} d^{\,(mn)}_{kl}\label{color}  \, ,
\end{align} 
 where the Greek letters refer to spin indices, the Roman letters refer to color indices, $\sigma^{i}$ denotes the Pauli matrices, ${\bm \epsilon}=i\sigma^2$ is an anti-symmetric 2 $\times 2$ matrix, and $d_{ij}^{\left(mn\right)}$ are symmetric matrices in color space:
\bea
d^{(mn)}_{ij} = \left\{ 
\begin{array}{cc}
(\delta^m_i \delta^n_j + \delta^n_i \delta^m_j)/\sqrt{2} & m\neq n \\
\delta^m_i \delta^n_j & m = n
\end{array} \right. \, .
\eea
 After Fourier transforming with respect to the labels and using the color and Fierz identities above, the Lagrangian in Eq.~\eqref{nrqcd} can be written as
\bea
{\cal L} &=&  
 -\frac{1}{4}F^{\mu\nu}F_{\mu \nu} + 
 \sum_{\bf p} \psip p ^\dagger   \Biggl ( i D^0 - \frac{\left({\bf p}-i{\bf D}\right)^2}{2 m_Q} 
 + \frac{g}{2 m_Q} \, \bm{\sigma}\cdot {{\bf B}} 
 \Biggr )
 \psip p  \\
&-&\frac{1}{2} \int d^3 {\bf r} \, V^{\left(\bm{\bar{3}}\right)}(r) \left( \sum_{\bf q} e^{-i \bf q \cdot r} \epsilon_{ijk}
\frac{1}{2}(\psi^\dagger_{\bf q})_j  \bm{\sigma} \epsilon (\psi^\dagger_{-\bf q})_k\right)\cdot
\left( \sum_{\bf p} e^{i \bf p \cdot r} 
\frac{1}{2} \epsilon_{ilm}(\psi_{\bf -p})_l \epsilon \bm{\sigma}(\psi_{\bf p})_m\right) \nn \\
&-&\frac{1}{2} \int d^3 {\bf r} \, V^{({\bm 6})}(r) \left( \sum_{\bf q} e^{-i \bf q \cdot r} 
\frac{1}{\sqrt{2}} d^{(mn)}_{ij} (\psi^\dagger_{\bf q})_i   \epsilon (\psi^\dagger_{\bf -q})_j\right) 
\left( \sum_{\bf p} e^{i \bf p \cdot r} 
\frac{1}{\sqrt{2}} d^{(mn)}_{kl} (\psi_{\bf -p})_k \epsilon^T (\psi_{\bf p})_l\right) \nn \, ,
\label{nrqcd_1}
\eea 
 where we have suppressed the spin indices and explicitly shown the color indices. The anti-triplet potential $V^{\left(\bm{\bar{3}}\right)}\left(r\right)$ and sextet potential $V^{\left({\bm 6}\right)}\left(r\right)$ are defined by 
\bea
V^{(\bm {\bar{3}})}(r) = -\frac{2}{3} \frac{\alpha_s}{r}, \qquad V^{({\bm 6})}(r) = \frac{1}{3} \frac{\alpha_s}{r} \, .
\eea

The color and spin Fierz identities in Eqs.~\eqref{spin} and \eqref{color} introduces four terms but two of them vanish due to Fermi statistics. The diquark fields in Eq.~\eqref{nrqcd_1} are in the $\bm{\bar{3}}$ and ${\bm 6}$ representations in color space, and have spin-1 and spin-0, respectively. We define the following composite diquark operators  
\begin{align}
 T_{\bm r}^{i}&=\sum_{\bf p} e^{i \bf p \cdot r} \frac{1}{2}\, \epsilon^{ijk} (\psi_{\bf -p})_j \epsilon \, \bm{\sigma} (\psi_{\bf p})_k,\label{spin1_diquark}\\
 \Sigma^{\left(mn\right)}_{\bm r}&= \sum_{\bf p} e^{i \bf p \cdot r} \frac{1}{\sqrt{2}} \, 
d^{(mn)}_{ij} (\psi_{\bf-p})_i \epsilon^T (\psi_{\bf p})_j  ,
\label{spin0_diquark}
\end{align}
 where $T_{\bm r}^{i}$ is a spin-1 vector field and $\Sigma^{\left(mn\right)}_{\bm r}$ is a spin-0 scalar field.

\begin{figure}[t]
	\centering
	\centerline{\includegraphics[width=10 cm,clip=true]{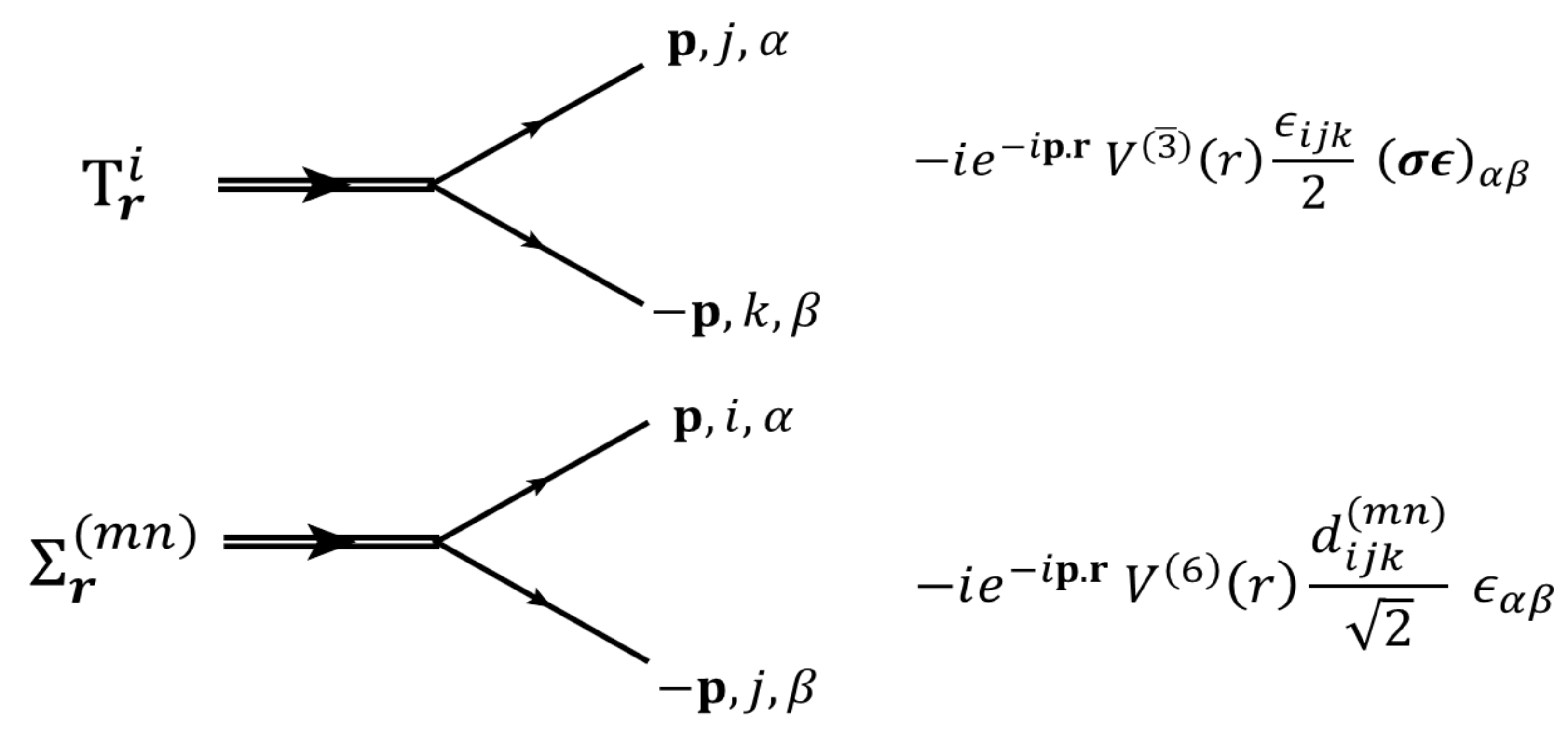} }
	\caption{Feynman rules for the coupling of the composite diquark fields to quarks.}
	\label{feynman_rules}
\end{figure}

The composite diquark fields, $T_{\bm r}^{i}$  and $\Sigma^{\left(mn\right)}_{\bm r}$, enter the theory by using the Hubbard–Stratonovich transformation, which cancels the quartic interaction terms in heavy quark fields in favor of interaction terms between the diquark fields and the two heavy quark fields:
\bea
\Delta {\cal L} &=& 
\frac{1}{2} \int d^3 {\bf r} \, V^{\left(\bm{\bar{3}}\right)}(r) \left({\bf T}^{i\dagger}_{\bf r} -  \sum_{\bf q} e^{-i \bf q\cdot r} \epsilon_{ijk}
\frac{1}{2}(\psi^\dagger_{\bf q})_j  \bm{\sigma} \epsilon (\psi^\dagger_{\bf -q})_k  \right) \nn \\
&&\qquad \times
\left({{\bf T}}^i_{\bf r}-\sum_{\bf p} e^{i \bf p\cdot r} 
\frac{1}{2} \epsilon_{ilm}(\psi_{\bf -p})_l \epsilon \bm{\sigma}(\psi_{\bf p})_m\right)  \nn \\
&& + \frac{1}{2} \int d^3 {\bf r} \, V^{({\bm  6})}(r) \left(\Sigma^{(mn)\dagger}_{\bf r} - \sum_{\bf q} e^{-i \bf q \cdot r} \frac{1}{\sqrt{2}} \,
d^{(mn)}_{ij} (\psi_{\bf q})_i \epsilon (\psi_{\bf -q})_j \right) \nn \\
&&\qquad \times \left( \Sigma^{(mn)}_{\bf r}- \sum_{\bf p} e^{i \bf p\cdot r} \frac{1}{\sqrt{2}} \,
d^{(mn)}_{ij} (\psi_{\bf-p})_i \epsilon^T (\psi_{\bf p})_j  \right) \, .
\label{DelL}
\eea
 The NRQCD Lagrangian after using the Hubbard-Stratonovich transformation reduces to 
\bea
{\cal L}+\Delta {\cal L} &=&  
 -\frac{1}{4}F^{\mu\nu}F_{\mu \nu} + 
 \sum_{\bf p} \psip p ^\dagger   \Biggl( i D^0 - \frac{\left({\bf p}-i{\bf D}\right)^2}{2 m_Q} 
 + \frac{g}{2 m_Q} \, \bm{\sigma}\cdot {{\bf B}} 
 \Biggr)\psip p\nn  \\
 &+&\frac{1}{2} \int d^3 {\bf r} \, V^{\left(\bm{\bar{3}}\right)}(r)\Biggl(
 {\bf T}^{i\dagger}_{\bf r}{\bf T}^i_{\bf r}-{\bf T}^{i\dagger}_{\bf r}\sum_{\bf p} e^{i \bf p\cdot r} 
\frac{1}{2} \epsilon_{ilm}(\psi_{\bf -p})_l \epsilon \bm{\sigma}(\psi_{\bf p})_m \nn\\
&&\hspace{4.2 cm}- \sum_{\bf q} e^{-i \bf q\cdot r} \epsilon_{ijk}\frac{1}{2}(\psi^\dagger_{\bf q})_j  \bm{\sigma} \epsilon (\psi^\dagger_{\bf -q})_k{\bf T}^i_{\bf r}\Biggr)\nn\\
&+& \frac{1}{2} \int d^3 {\bf r} \, V^{({\bm  6})}(r)
\Biggl(\Sigma^{(mn)\dagger}_{\bf r} \Sigma^{(mn)}_{\bf r}-
\Sigma^{(mn)\dagger}_{\bf r}\sum_{\bf p} e^{i \bf p\cdot r} \frac{1}{\sqrt{2}} \, d^{(mn)}_{ij} (\psi_{\bf-p})_i \epsilon^T (\psi_{\bf p})_j\nn\\
&&\hspace{4.2 cm}-\sum_{\bf q} e^{-i \bf q \cdot r} \frac{1}{\sqrt{2}} \,
d^{(mn)}_{ij} (\psi_{\bf q})_i \epsilon (\psi_{\bf -q})_j\Sigma^{(mn)}_{\bf r}\Biggr).
\label{Lagrangian:HS}
 \eea
The Feynman rules describing the interaction of diquarks with two heavy quarks corresponding to the above Lagrangian are shown in Fig.\ref{feynman_rules}. 

\begin{figure}[h!]
	\centering
	\centerline{ \includegraphics[width=10 cm,clip=true]{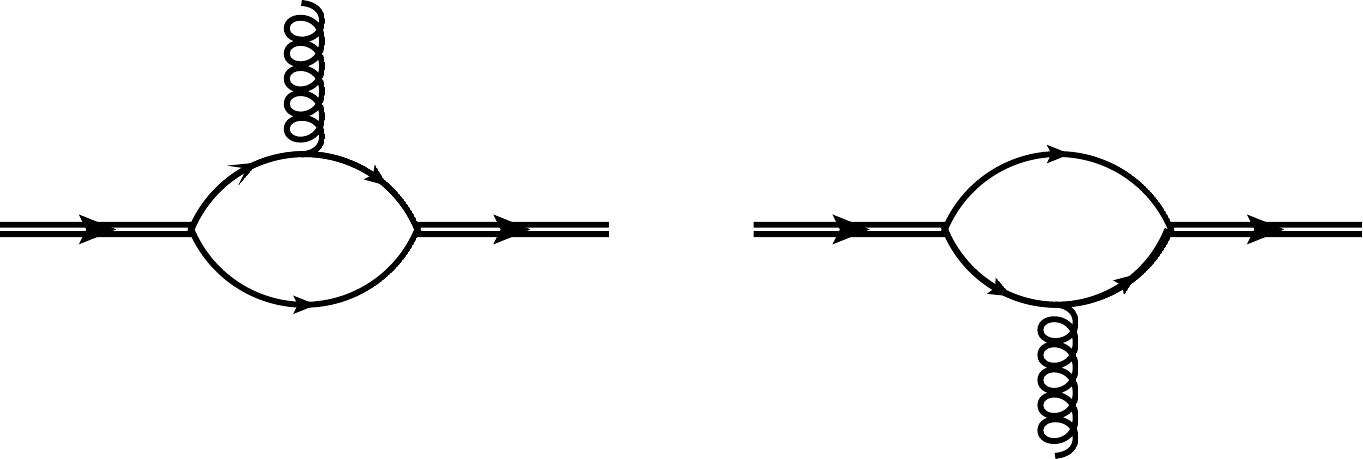} }
    \vspace*{0.0cm}
	\caption{Two one-loop diagrams at ${\cal O}\left(v^2\right)$ in NRQCD power counting that are responsible for the chromomagnetic coupling of diquarks.}
	\label{one_loop}
\end{figure}

The ${\bm \sigma}\cdot{\bm B}$ term in the NRQCD Lagrangian in Eq.~\eqref{nrqcd} is the chromomagnetic interaction for heavy quarks. This  is the lowest order heavy quark spin symmetry violating term that gives ${\cal O}\left(1/m_Q\right)$ corrections to the heavy quark spin symmetry and is responsible for the hyperfine splittings in the ground state of heavy mesons.
The chromomagnetic coupling for the heavy diquark field $T_{\bm r}^{i}$ was derived in  Ref.~\cite{Mehen:2006} by considering the two one-loop diagrams shown in Fig.~\ref{one_loop}, which contributes at ${\cal O}\left(v^2\right)$ to the effective action in the NRQCD power counting. The effective Lagrangian for the diquark field $T_{\bm r}^{i}$ which gives ${\cal O}\left(1/m_Q\right)$ corrections to the heavy quark spin symmetry and is responsible for the hyperfine splittings in the ground state of doubly heavy baryons is
\begin{equation}
{\cal L}_{\sigma.{\bm B}} = i \, \frac{g}{2 m_Q}\int d^3 {\bf r} \, {{\bf T}}^{i\,\dagger}_{\bf r} \cdot{\bf B}^c \,\bar{T}^c_{ij} \times{\bf T}^j_{\bf r}.
\label{chromo_coupling}
\end{equation}
Other ${\cal O}\left(v^2\right)$ couplings of the diquark field, $T_{\bm r}^{i}$, which do not violate the heavy quark spin symmetry can be found in Ref.~\cite{Brambilla:1999}.
 The composite diquark field, $\Sigma_{\bm r}^{\left(mn\right)}$, is a scalar and therefore does not have a chromomagnetic coupling. Other chromomagnetic couplings are possible if one considers diquarks composed of two different heavy quarks.

\begin{figure}[h!]
	\centering
	\centerline{ \includegraphics[width=10 cm,clip=true]{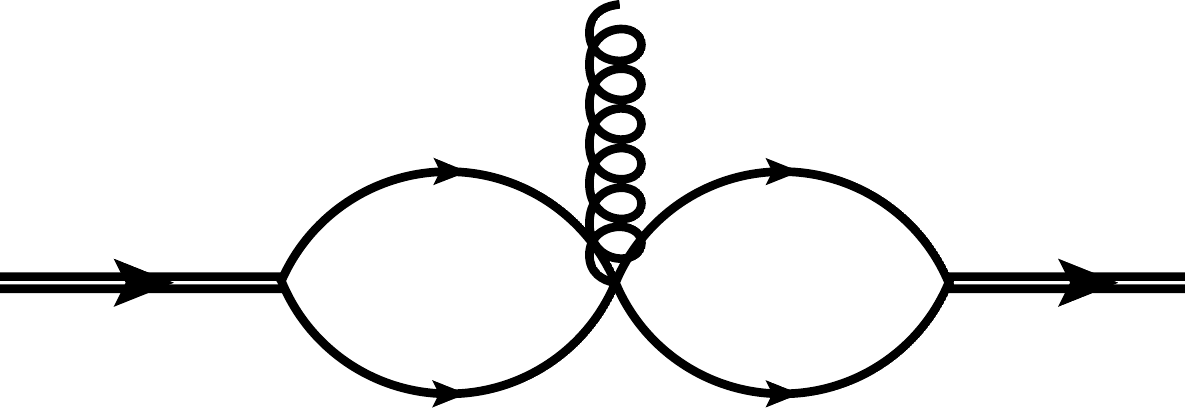} }
	\vspace*{0.0cm}
	\caption{The Two-loop diagram that give corrections to the chromomagnetic coupling of diquarks and contributes at ${\cal O}\left(v^4\right)$ to the effective action in NRQCD power counting.}
	\label{two_loop}
\end{figure}

\begin{figure}[h!]
	\centering
	\centerline{ \includegraphics[width=6 cm,clip=true]{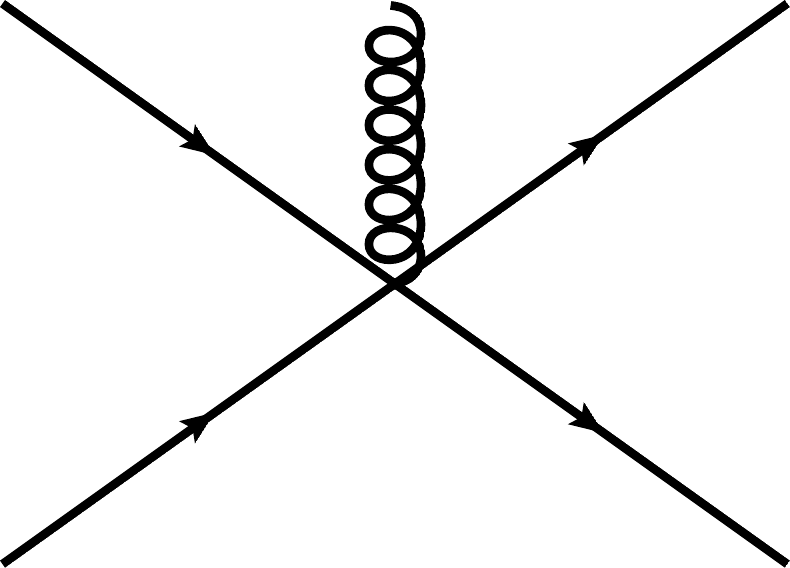} }
	\caption{Effective five point contact interaction with four heavy quarks and a gluon. This interaction gives ${\cal O}\left(1/m_Q^2\right)$ chromomagnetic coupling of diquarks.}
	\label{effective_operator}
\end{figure}

The effective Lagrangian for the  diquark field, $T_{\bm r}^{i}$, in Eq.~\eqref{chromo_coupling} can have corrections from terms that contribute at ${\cal O}\left(v^3\right)$ and higher to the effective action in the NRQCD power counting. The leading corrections to the chromomagnetic coupling of diquarks come from a two-loop diagram shown in Fig.~\ref{two_loop}.
The two-loop diagram contributes at ${\cal O}\left(v^4\right)$ to the effective action and gives ${\cal O}\left(1/m_Q^2\right)$ corrections to the heavy quark spin symmetry. It arises from an effective five point contact interaction shown in Fig.~\ref{effective_operator}, which is obtained after matching  tree-level scattering of two heavy quarks  and gluon in QCD and NRQCD. The Lagrangian for the effective operator in Fig.~\ref{effective_operator}  is given by Eq.~\eqref{effective_operator_Lagrangian} and
a detailed derivation is shown in Appendix~\ref{sec:operator}.

In order to evaluate the correction to the chromomagentic coupling of a diquark from the two-loop diagram in Fig.~\ref{two_loop}, we consider the external diquark fields $T_{\bm r}^{i}$ and $T_{\bm r^{'}}^{i}$ to be at rest and have energy $E$ and $E^{'}$ respectively. The external diquarks have spin indices $k$ and $l$  and color indices $a$ and $b$ respectively. The usoft gluon has polarization index $m$ and color index $c$. Using the Feynman rules for the diquark-quark interaction in Fig.~\ref{feynman_rules} and the effective four-quark contact vertex in Eq.~\eqref{effective_operator_Lagrangian}, the two-loop diagram in Fig.~\ref{two_loop} evaluates to
\begin{equation}
i\Sigma=-\frac{g^3}{6m_Q^2}\epsilon_{klm}\bar{T}^{c}_{ba}\int\frac{d^3{\bm l}}{\left(2\pi\right)^3}\,\int\frac{d^3{\bm l^{'}}}{\left(2\pi\right)^3}\,\frac{e^{-i{\bm l}\cdot{\bm r}}V^{\left({\bm \bar{3}}\right)}\left(r\right)}{E-l^2/m_Q+i\epsilon}\frac{e^{i{\bm l^{'}}\cdot{\bm r^{'}}}V^{\left({\bm \bar{3}}\right)}\left(r^{'}\right)}{E^{'}-l^{'2}/m_Q+i\epsilon}\,\frac{B^{c}_{m}}{|l-l^{'}|^2}.
\label{eq:two_loop}
\end{equation}
 The effective Lagrangian describing the leading correction to the chromomagnetic coupling of diquark field $T_{\bm r}$ in Eq.~\eqref{chromo_coupling} is
\begin{equation}
    {\cal L}^{'}_{\sigma.{\bm B}}= \int d^3 {\bf r}\,\int d^3 {\bf r^{'}}{\bf T}^{\dagger}_{\bf r}\,\Sigma\,{\bf T}_{\bf r^{'}},
    \label{eq:correction}
\end{equation}
 where $i\Sigma$ is given in Eq.~\eqref{eq:two_loop} and the color and spin indices of the diquark field, ${\bf T}_{\bf r}$, have been suppressed. This Lagrangian can be easily interpreted using the notation of Ref.~\cite{Mehen:2006}, where the diquark field ${\bf T}_{\bf r}$ are thought of as vectors in a Hilbert space spanned by the position space eigenkets $|{\bm r}\rangle$:
\begin{equation}
    {\cal L}^{'}_{\sigma.{\bm B}}=i \,\frac{g^3}{6m_Q^2}\epsilon_{klm}\bar{T}^c_{ba}B^c_m\,\langle T|\hat{V}^{({\bm \bar{3}})}\,\frac{1}{E^{'}-H_0^{'}}\,\frac{1}{4\pi r}\,\frac{1}{E-H_0}\,\hat{V}^{({\bm \bar{3}})}|T\rangle,
    \label{eq:correction_1}
\end{equation}       
where we define the diquark field $T_{\bm r}\equiv\langle {\bm r}|T\rangle$, the  potential operator $\langle {\bm r^{'}}|\hat{V}^{({\bm \bar{3}})}|{\bm r}\rangle \equiv V^{({\bm \bar{3}})}\delta^3\left({\bm r}-{\bm r}^{'}\right)$, the momentum eigenstates $\langle {\bm r}|{\bm l}\rangle=e^{-i{\bm l}\cdot{\bm r}}$, and a free Hamiltonian $H_0|{\bm l}\rangle=l^2/m_Q|{\bm l}\rangle$. Using the formalism developed in Ref.~\cite{Mehen:2006}, the potential operator $\hat{V}^{({\bm \bar{3}})}$ in Eq.~\eqref{eq:correction_1} cancels against the factors of $E-H_0$ in the denominator after using the equation of motion for the diquark field $T_{\bm r}$. Therefore, the effective Lagrangian describing the leading correction to the chromomagnetic coupling of diquarks in Eq.\eqref{chromo_coupling} is
\begin{equation}
    {\cal L}^{'}_{\sigma.{\bm B}}=i \, \frac{g}{2m_Q} \,\frac{\alpha_s}{3 m_Q}\int d^3 {\bf r} \, {{\bf T}}^{i\,\dagger}_{\bf r} \cdot \frac{1}{r}\,{\bf B}^c \, \bar{T}^c_{ij} \times{\bf T}^j_{\bf r}.
    \label{chromo_correction}
\end{equation}
 The effective Lagrangian in above equation contributes at ${\cal O}\left(v^4\right)$ to the effective action of diquarks because  the chromomagnetic field ${\bm B}$ scales as ${\cal O}\left(v^4\right)$, the diquark field $T_{\bm r}$ scales as ${\cal O}\left(v^3\right)$, the strong coupling constant $\alpha_s$ scales as ${\cal O}\left(v\right)$, the position vector ${\bm r}$ scales as ${\cal O}\left(v^{-1}\right)$, and the integration measure $d^4x$ scales as ${\cal O}\left(v^{-5}\right)$ in the NRQCD power counting that was developed in Ref.~\cite{Luke:1999kz}.

The mesons $P$ and $P^*$ and the baryons $\Xi$ and $\Xi^*$  are both degenerate in the absence of the chromomagnetic interactions that is suppressed by an inverse power of $m_Q$. The chromomagnetic coupling of heavy quarks in Eq.~\eqref{nrqcd} and the chromomagnetic coupling of heavy diquarks in Eqs.~\eqref{chromo_coupling} and \eqref{chromo_correction} lead to mass splittings of heavy mesons and doubly heavy baryons in ground state. Since the ground state of diquarks in the double heavy baryons is an s-wave $\left(l=0\right)$, the spatial wavefunction of the diquark can be approximated by a hydrogen-like spatial wavefunction in the limit of extremely large $m_Q$:
\begin{equation}
\phi\left({\bm r}\right)=\left(\frac{1}{\pi a_0^3}\right)^{1/2}e^{-r/a_0},
\label{eq:diquark_wavefunction}
\end{equation}
 where $a_0$ is the Bohr radius given by 
\begin{equation}
    a_0=\frac{3}{\alpha_s m_Q},
    \label{eq:Bohr}
\end{equation}

The identical configurations of light degrees of freedom in the heavy $\bar{Q}q$ meson and the heavy $QQq$ baryons 
implies that the hyperfine mass splittings in the heavy antimeson and the doubly heavy baryon are related by the heavy quark-diquark symmetry. The hyperfine splittings depends on the matrix elements of the chromomagnetic couplings of quarks and diquarks in Eqs.~\eqref{nrqcd}, \eqref{chromo_coupling}, and \eqref{chromo_correction}. The matrix element of the ${\cal O}\left(1/m_Q^2\right)$ chromomagnetic coupling of diquark in Eq.~\eqref{chromo_correction} depends on $\langle 1/r\rangle$:
\begin{equation}
  \left\langle\frac{1}{r}\right\rangle=\int d^3{\bm r}\, \frac{|\phi\left({\bm r}\right)|^2}{r}=\frac{1}{a_0}.
  \label{eq:expectation}
\end{equation}
 This gives a correction to the matrix element of the chromomagnetic operator in Eq.~\eqref{chromo_coupling} that appears only in the doubly heavy baryon mass splitting as there is no analogue correction in the heavy meson sector. Therefore, the relation between the hyperfine splittings in Eq.~\eqref{hyperfine:HQDQ} is modified to
\begin{align}
    m_{P^*}-m_P&=\frac{4}{3}\left(m_{\Xi^*}-m_{\Xi}\right)\left(1+\frac{\alpha_s}{3m_Q}\left\langle\frac{1}{r}\right\rangle\right),\nonumber\\
    &=\frac{4}{3}\left(m_{\Xi^*}-m_{\Xi}\right)\left(1+\frac{\alpha_s^2}{9}\right),
    \label{eq:hyperfine}
\end{align}
 where $\alpha_s^2/9$ is the correction to the hyperfine splitting from the two-loop diagram in Fig.~\ref{two_loop}. Of the two powers of $\alpha_s$ appearing in Eq.~(\ref{eq:hyperfine}), one arises from matching QCD onto NRQCD, so naturally lives at the scale $m_Q$, while the other appears in the evaluation of the matrix element, $\langle1/r\rangle$, and naturally lives at the scale $m_Q v$. It would be interesting to compute the anomalous dimension of the operator in Eq.~\eqref{chromo_correction} in order to sum logarithms of ratios of these two scales, but that is beyond the scope of this work.

For numerical purposes, we will use the value of $\alpha_s$ evaluated at the scale $m_Qv$. If the doubly heavy baryons are composed of charm quarks, the value of the strong coupling constant is $\alpha_s\left(m_cv\right)\approx0.52$, which implies the correction to the hyperfine splitting  is $3\times10^{-2}$. If the doubly heavy baryons are composed of bottom quarks, the value of strong coupling constant is $\alpha_s\left(m_bv\right)\approx0.35$, which implies the correction to the hyperfine splitting  is $1.4\times10^{-2}$. Therefore, we conclude the heavy quark diquark symmetry prediction receives very small correction at ${\cal O}\left(v^2\right)$, at least as $m_Q\rightarrow\infty$.

\section{Nonperturbative Corrections to Hyperfine Splittings} 

Having established that the perturbative corrections to the HQDQ symmetry prediction for the hyperfine splitting are small, we now consider nonperturbative corrections which scale as powers of $\Lambda_{\rm QCD}/m_Q$.\footnote{We thank Nora Brambilla for comments on an earlier version of this paper that led us to add this section.}  We point out that the spin-color structure of the operator that contributes to the hyperfine splittings at $O(1/m_Q^2)$ is the same as the leading operator.  Therefore, its corrections to Eq.~(\ref{hyperfine:HQDQ}) should give contributions that are  consistent with the relative factor of 3/4 between the hyperfine splittings.   Thus we expect the corrections to Eq.~(\ref{hyperfine:HQDQ}) will
be $O(\Lambda_{\rm QCD}^2/m_Q^2)$, which will be $\approx 10\%$ for charm and smaller for bottom. 

The hyperfine splittings in Eq.~(\ref{hyperfine:HQDQ}) vanish in the absence of heavy quark spin symmetry violation. So, we only need to consider the heavy quark spin symmetry violating operators in the HQET Lagrangian. To $O(\Lambda_{\rm QCD}^2/m_Q^2)$ these are
\bea
\frac{g}{2 m_Q} \sum_{\bf p}\psi_{\bf p}^\dagger \, \bm{\sigma} \cdot {\bf B} \, \psi_{\bf p} + i\frac{g}{8 m_Q^2}\sum_{\bf p} \psi^\dagger_{\bf p} \, \bm{ \sigma} \cdot ({\bf D\times E - E \times D}) \, \psi_{\bf p}  \,.
\eea
Both operators have the same color and spin structure, in the second operator the heavy quark spin couples to a different background field:
${\bf B} \to i({\bf D\times E -   E \times D})/(4 m_Q)$. Thus by a calculation that is essentially identical to the one in Ref.~\cite{Mehen:2006} one finds the corresponding Lagrangian for diquarks:
\begin{equation}
    {\cal L} =i \, \frac{g}{2m_Q}  \int d^3 {\bf r} \, {{\bf T}}^{i\,\dagger}_{\bf r} \cdot \,{\bf B}^c \, \bar{T}^c_{ij} \times{\bf T}^j_{\bf r} -
    \frac{g}{8m_Q^2} \int d^3 {\bf r} \, {{\bf T}}^{i\,\dagger}_{\bf r} \cdot \,({\bf D\times E}^c - {\bf D\times E}^c )\, \bar{T}^c_{ij} \times{\bf T}^j_{\bf r}.
    \label{TLagatmsquared}
\end{equation}
The arguments in Ref.~\cite{Savage:1990di} that lead to the factor of 3/4 relating the two hyperfine splitting hold for both operators in Eq.~(\ref{TLagatmsquared}). Note that in the quark model the  factor ${\bf D\times E} - {\bf D\times E}$ leads to spin-orbit couplings that vanish in the ground state meson and doubly heavy baryons because all constituents are in an $S$-wave. In full QCD, there could be a nonvanishing contribution but the group theoretical arguments that lead to the factor of 3/4 still apply. 

At $O(1/m_Q^3)$ in the HQET Lagrangian there are 11 operators, several of which violate heavy quark spin symmetry. These have different color structures than the leading two operators so we expect that these operators will give corrections to Eq.~(\ref{hyperfine:HQDQ}) at $O(\Lambda_{\rm QCD}^2/m_Q^2)$.

\section{Conclusion}
The ground state mass hyperfine splittings in the double heavy baryons and singly heavy antimesons are related by the heavy quark-diquark (HQDQ) symmetry. The hyperfine splittings are due to the ${\cal O}\left(1/m_Q\right)$ chromomagnetic couplings of the diquark and quark and the leading prediction for the splittings is given by Eq.~\eqref{hyperfine:HQDQ}. In this paper, we compute the leading correction to the hyperfine splitting of the double heavy baryons in the framework of NRQCD.  We point out an effective five-point contact operator that couples the four heavy quark fields with the chromomagnetic field with a coefficient that is enhanced by the Coulomb exchange. Naively, one would expect the leading correction to the chromomagnetic coupling of the diquark to scale as 
$1/m_Q^2$, instead we find that the correction from the effective operator scales as $1/m_Q^2 \times \alpha_s/r$, where $r$ is the separation between the heavy quarks in the diquark. The Lagrangian describing the leading correction to the chromomagnetic coupling of diquark is given by Eq.~\eqref{chromo_correction}. 

We estimate the correction to the ground state mass hyperfine splitting in the doubly heavy baryons due to the next leading order Lagrangian in Eq.~\eqref{chromo_correction}. We find that in the $m_Q\rightarrow\infty$ limit, when the two quarks within the diquark are bound by  strong Coulombic interaction, the leading correction to the hyperfine splitting of double heavy baryons is of ${\cal O}\left(\alpha_s^2/m_Q\right)$ with a small coefficient as shown in Eq.~\eqref{eq:hyperfine}. This ${\cal O}\left(\alpha_s^2\right)$ correction to the hyperfine splittings was anticipated in Ref.~\cite{Brambilla:2005} but we have explicitly calculated the correction in this paper. For values of $\alpha_s$ relevant to doubly charm and doubly bottom systems, we find that the correction to the hyperfine splitting in doubly heavy baryons is $3\times10^{-2}$ for doubly charm baryons and $1.4\times10^{-2}$ for doubly bottom baryons. 
We also gave an argument why corrections
to Eq.~(\ref{hyperfine:HQDQ})  should scale as $O(\Lambda_{\rm QCD}^2/m_Q^2)$. Therefore, we expect nonperturbative corrections to  Eq.~(\ref{hyperfine:HQDQ}) to be 10\% for charm and smaller for bottom. This is consistent with lattice calculations of doubly charm spectra \cite{Lewis:2001, Lewis:2002, Flynn:2003, Chiu:2005zc, Na:2007, Na:2008, Briceno:2012wt, Namekawa:2013vu, Alexandrou:2014sha, Brown:2014, Padmanath:2015jea, Bali:2015, Chen:2017, Alexandrou:2017, Padmanath:2019}.

\begin{acknowledgments}

This research is supported in part by  Director, Office of Science, Office of Nuclear Physics,
of the U.S. Department of Energy under grant number
DE-FG02-05ER41368. We thank M.B. Wise and N. Brambilla for useful discussions.

 \end{acknowledgments}
\appendix
\section{Effective five-point contact
operator}\label{sec:operator}
\begin{figure}[h!]
	\centering
	\centerline{ \includegraphics[width=10 cm,clip=true]{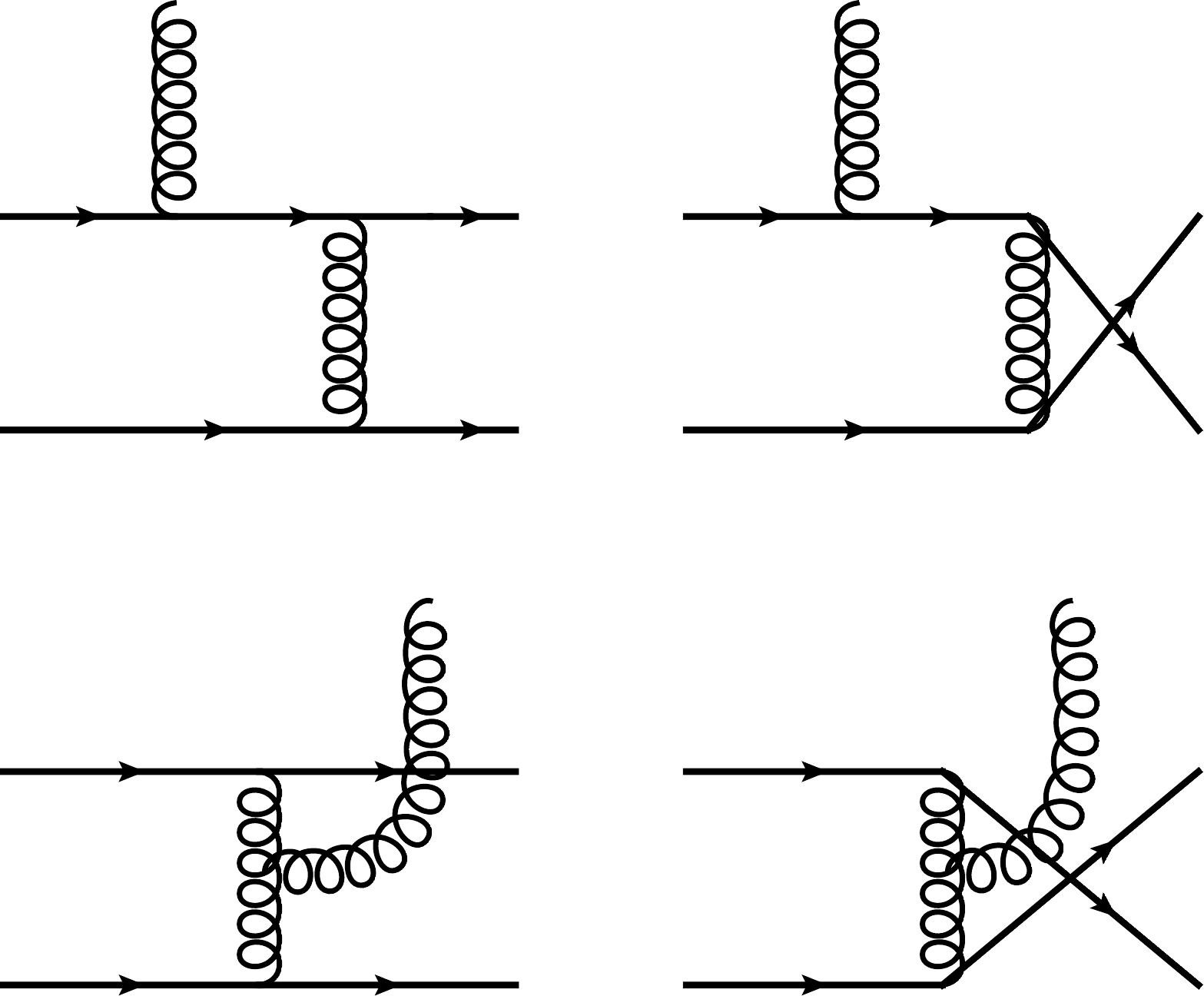} }
	\caption{Full QCD diagrams for $QQ
	\to QQg$. There are six other diagrams in QCD similar to the upper two diagrams, in which the external gluon is attached to each of the external quarks. 
	}
	\label{QCD_diagram}
\end{figure}

\begin{figure}[h!]
	\centering
	\centerline{\includegraphics[width=5 cm,clip=true]{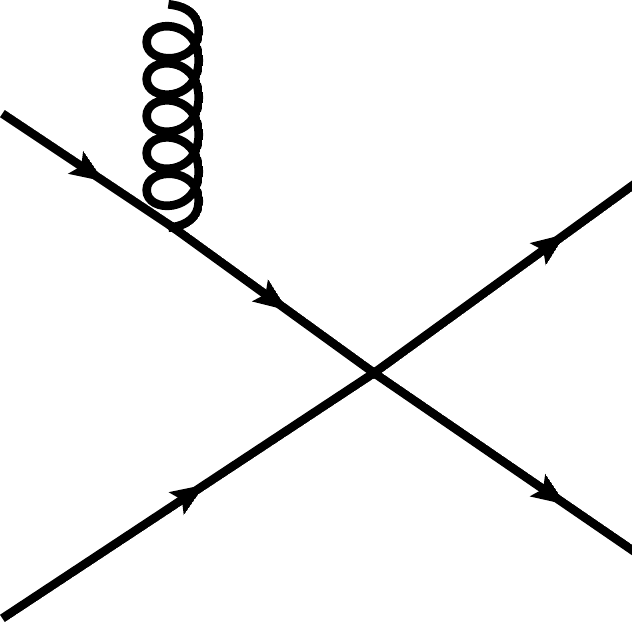} }
	\vspace*{0.0cm}
	\caption{One of the two types of NRQCD required to reproduce the full QCD result at low energy. The other is the contact interaction shown in  Fig.~\ref{effective_operator}. 
	}
	\label{NRQCD_diagram}
\end{figure}

In this Appendix, we derive the Lagrangian for the effective contact operator shown in Fig.~\ref{effective_operator}.
The effective five-point contact operator with four heavy quarks and one gluon in Fig.~\ref{effective_operator} gives an ${\cal O}\left(1/m_Q^2\times\alpha_s/r\right)$ correction to the chromomagnetic coupling of diquark field $T_{\bm r}$. This effective operator is obtained after matching the low-energy tree diagrams for $QQ\to QQg$ in full QCD theory onto NRQCD. In QCD, the diagrams for $QQ\to QQg$ are shown in Fig.~\ref{QCD_diagram}.  The two tree-diagrams at the top of Fig.~\ref{QCD_diagram}, where the external gluon is attached to the external quarks, match onto two distinct types of NRQCD diagrams. One diagram is the tree diagram shown in Fig.~\ref{NRQCD_diagram}, in which the gluon couples to an external quark via the chromomagnetic interaction and there is a virtual nonrelativistic quark. The other diagram is the contact interaction in Fig.~\ref{effective_operator}. The bottom two diagrams in Fig.~\ref{QCD_diagram}, where the external gluon is attached to the exchanged gluon via the three-gluon vertex, could in principle also contribute.  However, the bottom two diagrams in Fig.~\ref{QCD_diagram} have a vanishing color factor when the incoming and outgoing diquarks are both in the ${\bm \bar{3}}$ representation.  

In Fig.~\ref{QCD_diagram}, the incoming heavy quarks have four-momenta $p_1=\left(E_1,{\bm P_1 }\right)$ and $p_2=\left(E_2,{\bm P_2}\right)$ and color indices $i$ and $j$ respectively. The outgoing heavy quarks have four momenta given by $p_3=\left(E_3,{\bm P_3}\right)$ and $p_4=\left(E_4,{\bm P_4}\right)$ and color indices $r$ and $l$ respectively. The external gluon have four-momenta $q$ and color index $c$. In full QCD, the contribution to the low-energy scattering amplitude from the upper two diagrams of Fig.~\ref{QCD_diagram} is 
\begin{equation}
i{\cal A}=ig^3 T^{a}_{rk^{'}}T^{c}_{k^{'}i}T^{a}_{lj}\varepsilon_m\frac{\left[\bar{u}_l\left(p_4\right)\gamma_{\mu}u_j\left(p_2\right)\right]\left[\bar{u}_r\left(p_3\right)\gamma^{\mu}\left(\slashed{k}+m_Q\right)\gamma^{m}u_i\left(p_1\right)\right]}{\left(k^2-m_Q^2+i\epsilon\right)\left(p^2+i\epsilon\right)}
-\left(p_3\longleftrightarrow p_4, r\longleftrightarrow l\right),
\label{eq:QCD_amplitude}
\end{equation}
 where $p=p_4-p_2$ is the four-momenta of the intermediate gluon in the first term, $k=p_3+p$ is the four-momenta of the intermediate fermion in the first term, $\varepsilon^m$ is the polarization 4-vector of the incoming gluon and $m_Q$ is the mass of the heavy quark. In the above expression we have explicitly shown the color indices and suppressed the spinor indices. Using the equation of motion for the external states,  the amplitude in Eq.~\eqref{eq:QCD_amplitude} can be written as
\begin{equation}
i{\cal A}=ig^3 T^{a}_{rk^{'}}T^{c}_{k^{'}i}T^{a}_{lj}\varepsilon_m\frac{\left[\bar{u}_l\left(p_4\right)\gamma_{\mu}u_j\left(p_2\right)\right]\left[\bar{u}_r\left(p_3\right)\left(2p_3^{\mu}+\gamma^{\mu}\slashed{p}\right)\gamma^{m}u_i\left(p_1\right)\right]}{\left(k^2-m_Q^2+i\epsilon\right)\left(p^2+i\epsilon\right)}-\left(p_3\longleftrightarrow p_4, r\longleftrightarrow l\right).
\label{eq:QCD_amplitude_1}
\end{equation}
 Using the identity for the product of three gamma matrices, the terms in the second square bracket in the numerator of the above equation can rewritten as 
\begin{equation}
\bar{u}\left(p_3\right)\left(2p_3^{\mu}\gamma^{m}\varepsilon_m+p^{\mu}\gamma^{m}\varepsilon_m-\varepsilon^{\mu}p_m\gamma^{m}+p^m\varepsilon_m\gamma^{\mu}-i\epsilon^{\sigma \mu\alpha m}\gamma_{\sigma}p_{\alpha}\varepsilon_m\gamma^5\right)u\left(p_1\right). 
\label{eq:numerator}
\end{equation}
In the above expression, $p_3\sim{\cal O}\left(m_Q\right)$ and $p=p_4-p_2\sim{\cal O}\left(m_Qv\right)$, thus, the leading contribution comes only from the first term in the parenthesis. The  low-energy scattering amplitude for $QQ\to QQg$ can be obtained from QCD  by doing the nonrelativistic expansion of the amplitude in Eq.~\eqref{eq:QCD_amplitude} in powers of 3-momenta. Using the Gordon identity
 \begin{equation}
 \bar{u}\left(p\right)\gamma^{\mu}u\left(q\right)=\bar u(p) \left[\frac{(p+q)^\mu}{2m} +i \sigma^{\mu\nu}\frac{(p-q)_\nu}{2m}\right]u(q),
 \label{eq:Gordon}
 \end{equation}
and then taking the nonrelativistic limit of the Dirac spinors
\begin{equation}
    u_i\left({\bm P}\right)=
    \begin{pmatrix}
     \xi_i\\
    0
    \end{pmatrix},
    \label{eq:expansion_spinors}
    \end{equation}
 where  $\xi_i$ is a two-spinor, and $i$ denotes the color index,
 we find that the leading term with a spin-dependent interaction
 in the nonrelativistic expansion of the numerator in Eq.~\eqref{eq:QCD_amplitude_1} is
 \begin{equation}
2p_3^0 \left[\bar{u}\left(p_4\right)\gamma_0 u\left(p_2\right)\right]\left[\bar{u}\left(p_3\right)\slashed{\epsilon} u\left(p_1\right)\right]=-i\xi_{4,l}^{\dagger}\,\xi_{2,j}\,\xi_{3,r}^{\dagger}\,{\bm \sigma}.\left({\bm q}\times{\bm\varepsilon}\right)\,\xi_{1,i}\,+\cdots,
 \label{eq:numerator_1}
\end{equation}
 where ${\bm\varepsilon}$ is the three-space polarization vector and the ellipsis represents terms that are suppressed by powers of $v$ or do not explicitly break the heavy quark spin symmetry.

In the nonrelativistic limit, the fermion propagator in the denominator of the amplitude in Eq.~\eqref{eq:QCD_amplitude_1} is given by
\begin{equation}
    \frac{1}{k^2-m_Q^2+i\epsilon}=\frac{1}{2m_Q}\Bigg[\frac{1}{k_0-E_{\bm k}+i\epsilon}-\frac{1}{2m_Q}\Bigg]+{\cal O}\left(P^2/m_Q^4\right),
    \label{eq:fermion_propagator}
\end{equation}
 where ${\bm k}={\bm P_3}+{\bm P_4}-{\bm P_2}$ and
\begin{equation}\label{momenta_expansion}
k_0-E_k=\frac{P_3^2}{2m_Q}+\frac{P_4^2}{2m_Q}-\frac{P_2^2}{2m_Q}-\frac{\left({\bm P_3}+{\bm P_4}-{\bm P_2}\right)^2}{2m_Q}
\end{equation}
 Similarly, the gluon propagator can also be expanded in powers of 3-momenta as
\begin{equation}
    \frac{1}{\left(p_4-p_2\right)^2+i\epsilon}=-\frac{1}{\left({\bm P_4}-{\bm P_2}\right)^2}\left(1+{\cal O}\left(P^2/m_Q^2\right)\right),
    \label{eq:gluon_expansion}
\end{equation}

Using the nonrelativistic expansion of the numerator in Eq.~\eqref{eq:numerator_1} and the nonrelativistic expansion of the fermion and gluon propagators in Eq.~\eqref{eq:fermion_propagator} and \eqref{eq:gluon_expansion}, the low-energy scattering amplitude in Eq.~\eqref{eq:QCD_amplitude_1} is 
\begin{align}
    i{\cal A}=ig^3T^{a}_{rk^{'}}T^{c}_{k^{'}i}T^{a}_{lj}\Bigg[\xi_4^{\dagger}\,\xi_2\,\xi_3^{\dagger}\frac{{\bm \sigma}\cdot{\bm B}^{c}}{2m_Q}\xi_1\Bigg]&\Bigg[\frac{1}{k_0-E_{\bm k}+i\epsilon}-\frac{1}{2m_Q}\Bigg]\frac{1}{\left({\bm P_4}-{\bm P_2}\right)^2}\nonumber\\
    &-\left(P_3\longleftrightarrow P_4, r\longleftrightarrow l\right)+\cdots,
    \label{eq:amplitude_expansion}
\end{align}
 where ${\bm k}={\bm P_3}+{\bm P_4}-{\bm P_2}$ and $k_0-E_k$ is given by Eq.~\eqref{momenta_expansion}.  In the above expression we have suppressed both the color and spin indices. The low-energy amplitude in Eq.~\eqref{eq:amplitude_expansion} has a pole where the intermediate fermion propagator goes on-shell. This pole will be reproduced in the effective theory 
 by the contribution to the scattering amplitude from Fig.~\ref{NRQCD_diagram}.   
 If the vertex with the external gluon line is the chromomagnetic coupling of the heavy quark, then the contribution to the scattering amplitude from Fig.~\ref{NRQCD_diagram} is 
\begin{align}
    i{\cal A}=ig^3T^{a}_{rk^{'}}T^{c}_{k^{'}i}T^{a}_{lj}\Bigg[\xi_4^{\dagger}\,\xi_2\,\xi_3^{\dagger}\frac{{\bm \sigma}\cdot{\bm B}^{c}}{2m_Q}\xi_1\Bigg]&\Bigg[\frac{1}{k_0-E_{\bm k}+i\epsilon}\Bigg]\frac{1}{\left({\bm P_4}-{\bm P_2}\right)^2}\nonumber\\
    &-\left(\xi_4\longleftrightarrow \xi_3, P_4\longleftrightarrow P_3, r\longleftrightarrow l\right),
    \label{eq:NRQCDamplitude}
\end{align}
 ${\bm k}={\bm P_3}+{\bm P_4}-{\bm P_2}$ and $k_0-E_k$ is given by Eq.~\eqref{momenta_expansion}. In the above expression we have suppressed both the color and spin indices.
On comparing Eqs.~\eqref{eq:amplitude_expansion} and \eqref{eq:NRQCDamplitude}, we see that the pole in the low-energy scattering amplitude in Eq.~\eqref{eq:amplitude_expansion} from the intermediate fermion propagator is cancelled exactly by the corresponding contribution in NRQCD. After cancelling the pole, we are left with a term which is reproduced by the effective four-quark and gluon contact interaction shown in Fig.~\ref{effective_operator}. The five-point operator with four heavy quarks and one gluon receives contributions from six other diagrams in QCD and three other diagrams in NRQCD where the external gluon is attached to each of the external quarks. After taking into consideration all these diagrams, the Lagrangian for the five-point contact interaction with four heavy quarks and one gluon in Fig.~\ref{effective_operator} is 
\begin{equation}
    {\cal L}_{\mathrm eff}=-\frac{g^3}{2}\frac{1}{2m_Q}\sum_{{\bm P_1},{\bm P_2},{\bm P_3},{\bm P_4}}\psi^{\dagger}_{\bm P_4}T^{a}\psi_{\bm P_2}\psi^{\dagger}_{\bm P_3}\left(T^{a}T^{c}+T^{c}T^{a}\right)\frac{{\bm \sigma}\cdot{\bm B}^{c}}{2m_Q}\psi_{\bm P_1}\frac{1}{\left({\bm P_4}-{\bm P_2}\right)^2}
    \label{effective_operator_Lagrangian}
\end{equation}
The Feynman rule for the contact vertex in Fig.~\ref{effective_operator} is given by
\begin{equation}
-i \frac{g^3}{4m_Q^2}T^a\otimes \left(T^{a}T^{c}+T^{c}T^{a}\right)
\frac{{\bm\sigma}\cdot{\bm B}^{c}}{\left({\bm P_4}-{\bm P_2}\right)^2}.
\end{equation}

\end{document}